\documentclass[aas2pp4]{aastex}

\begin{document}


\title{Gravitational Bar and Spiral Arm Torques from $K_{\rm s}$-band
Observations and Implications for the Pattern Speeds} 


\author{D. L. Block\altaffilmark{1}, R. Buta\altaffilmark{2}, J. H. Knapen\altaffilmark{3}, D. M. Elmegreen\altaffilmark{4}, B. G. Elmegreen\altaffilmark{5}, \& I. Puerari\altaffilmark{6}}
\affil{}

%\email{aastex-help@aas.org}


\altaffiltext{1}{School of Computational and Applied Mathematics, University of the Witwatersrand,
P.O. Box 60, Wits, Gauteng 2050, South Africa}
\altaffiltext{2}{Department of Physics \& Astronomy, University of Alabama, Box 870324, Tuscaloosa, AL 35487}
\altaffiltext{3}{Centre for Astrophysics Research, Science \& Technology Research School, University of Hertfordshire, Hatfield, Herts AL10 9AB, U.K.}
\altaffiltext{4}{Department of Physics \& Astronomy, Vassar College, Box 745, Poughkeepsie, NY 12604}
\altaffiltext{5}{IBM Research Division, T. J. Watson Research Center, P. O. Box 218, Yorktown Heights,
NY 10598}
\altaffiltext{6}{Instituto Nacional de Astrof\'\i sica, Optica y Electr\'onica, Apdo, Postal 51 y 216, 72000, Puebla,
Pue., M\'exico}


\begin{abstract}

We have obtained deep near-infrared $K_{\rm s}$-band William
Herschel Telescope observations of a sample of 15 nearby spiral
galaxies having a range of Hubble types and apparent bar
strengths. The near-infrared light distributions are converted into
gravitational potentials, and the maximum relative gravitational
torques due to the bars and the spirals are estimated.  We find
that spiral strength, $Q_{\rm s}$, and bar strength, $Q_{\rm b}$,
correlate well with other measures of spiral arm and bar
amplitudes, and that spiral and bar strengths also correlate well
with each other.  We also find a correlation between the position
angle of the end of the bar and the position angle of the inner
spiral.  These correlations suggest that the bars and spirals grow
together with the same rates and pattern speeds. We also show that
the strongest bars tend to have the most open spiral patterns.
Because open spirals imply high disk-to-halo mass ratios, bars and
spirals most likely grow together as a combined disk instability.
They stop growing for different reasons, however, giving the
observed variation in bar-spiral morphologies. Bar growth stops
because of saturation when most of the inner disk is in the bar,
and spiral growth stops because of increased stability as the gas
leaves and the outer disk heats up.

\end{abstract}


\keywords{galaxies: spiral;  galaxies: photometry; galaxies: kinematics
and dynamics; galaxies: structure}


\section{Introduction}

The evolution of disk galaxies is significantly influenced by the
main features within their disks, notably spirals and bars. In
turn, galactic evolution also significantly influences the
evolution, morphology, and dynamics of these features. Bars at the
present epoch can be found in about 70\% of massive disk galaxies
using near-infrared images (Sellwood \& Wilkinson 1993; Knapen et
al. 1999; Eskridge et al. 2000). The fraction is somewhat lower on
optical images.

The gravitational torque method (GTM, Buta \& Block 2001) was
developed to quantify bar strength. The idea is to transform a
deprojected near-infrared image of a spiral galaxy into a
gravitational potential, and then compute the ratio of the
tangential force to the mean radial force as a function of
position in the plane of the galaxy. The potential is derived from
Poisson's law after assuming an exponential vertical density law
and a constant mass-to-light ratio (Quillen, Frogel, and
Gonz\'alez 1994). The mean radial force represents the
axisymmetric background due to the bulge, disk, and bar. A fully
quantitative measure of bar strength can be defined from the
maximum of the tangential-to-radial force ratio, as long ago
suggested by Sanders \& Tubbs (1980) and Combes \& Sanders (1981).
The maximum force ratio is equivalent to the maximum gravitational
torque per unit mass per unit square of the circular speed. The
GTM has been applied to large samples of galaxies in several
recent studies (Block et al. 2001, 2002; Laurikainen, Salo, \&
Rautiainen 2002; Laurikainen \& Salo 2002; Buta, Laurikainen, \&
Salo 2004, hereafter BLS; Laurikianen, Salo, \& Buta 2004a;
Laurikainen et al. 2004b). In general, the GTM has an advantage
over other methods previously used to define bar strength because
it is based on the forcing due to the bar itself and not just on
the bar's apparent shape. The relative bar torque is a good
measure of the importance of non-axisymmetric forces, which play a
role in gas accretion, spiral arm generation, and overall disk
evolution.

Studies at optical wavelengths suggested that early-type barred
galaxies are associated with grand design spiral structure because
the bars end near their own corotation resonances, and at this
point they have enough torque to drive corotating spiral waves
outward into the disk (e.g., Elmegreen \& Elmegreen 1985, 1989).
Late-type bars could have different bar and
spiral pattern speeds, as observed for NGC 925 (Elmegreen, Wilcots
\& Pisano 1998).

The interaction between bars and spirals is not well understood
observationally because most studies have
been based on optical images, which are
confused by dust and star formation.
We have shown, for example, that $K$-band spirals
often look different from optical spirals with regard to
pitch angle, continuity,
and symmetry (Block \& Wainscoat 1991; Block et al. 1999). Because
the light distribution at 2.2$\mu$m emphasizes the mass distribution
in the old disk, the near-infrared spirals are more important
for bar driving than the optical spirals.

Buta, Block, \& Knapen (2003, hereafter BBK) have shown that it is
possible to use straightforward Fourier techniques to separate the
luminosity distributions of bars and spirals (see also Lindblad,
Lindblad, \& Athanassoula 1996). The GTM is then used to derive
maximum relative torques for each nonaxisymmetric feature alone,
thereby allowing us to examine in a quantitative manner the
possible correlation between bar and spiral strengths. In this
paper, we apply the BBK method to 17 representative disk galaxies
covering a range of early to intermediate Hubble types and de
Vaucouleurs family classifications.  Our goal is to investigate
the relation between bar torque and spiral arm amplitude.

\section{Observations and Sample}

Our sample consists of the 17 galaxies listed in Table 1. These
were selected on the basis of Hubble type (range Sab-Scd) and
inclination (less than 70$^{\circ}$). The absolute blue magnitudes
of the galaxies range from $-$18.3 to $-$21.4, with an average of
$-$20.5$\pm$0.9. Thus, these are massive, high luminosity systems,
typical of the bright galaxy population. A range of apparent bar
strengths covering the de Vaucouleurs families SA, SAB, and SB was
also selected.

Fifteen of our sample galaxies were observed during a total of five
nights (2001 September 5, and 2001 October 5--8) with the INGRID
camera (Packham et al. 2003) attached to the 4.2-m William Herschel Telescope
(WHT). The images were taken in the $K$-short (2.2$\mu$m, or $K_{\rm s}$)
band, and have a scale of 0.241~arcsec/pix and a field of view approximately 4
arcminutes square. Total on-source exposure times average about 59 min,
but ranged from 16 to 100 min (see Table 1).  The observing techniques are
the same as those described in Knapen et al. (2003). Particularly important
is the background subtraction, which was performed by interspersing between
small blocks of galaxy observations several exposures of the same length
on a blank background field. We estimate the typical sky subtraction error
to be on the order of 0.1\%-0.2\% of the total background.
To eliminate bad pixels, columns, and (in
the case of the background frames) foreground stars, dithering was used
for both the galaxy and background frames. We
required deeper than usual exposures at 2$\mu$m in order to effectively measure
spiral arm torques. The limiting surface brightness and typical signal-to-noise
ratio were estimated from our $K_{\rm s}$ image of NGC 1530 using an independently
measured $K_{\rm s}$ flux listed in the NED. In this image, a 48 min
exposure, the signal-to-noise ratio is 2.6 at a surface brightness level
$\mu_{K_{\rm s}}$ = 20.0 mag arcsec$^{-2}$.
Typical $1\sigma$ background noise is 21.0 mag\,arcsec$^{-2}$, down
to around 21.3 mag\,arcsec$^{-2}$ for our longest exposures.

In addition to the 15 WHT objects, our analysis includes
two southern strongly-barred galaxies, NGC~1365 and NGC~1433, which were part
of previously published studies. These
were observed with the CTIO Infrared Imager (CIRIM) attached to the
1.5-m telescope of Cerro Tololo Inter-American Observatory (CTIO) and
have a scale of 1\rlap{.}$^{\prime\prime}$14 pix$^{-1}$.
Details of the observations of NGC~1365, observed at $K$, are provided
by Regan \& Elmegreen (1997), while those for NGC~1433, observed at $H$,
are provided by Buta et al. (2001).

\section{Bar/Spiral Separation}

The GTM as used in previous studies involves the derivation of the
maximum tangential-to-radial force ratio, a single number that characterizes
the strength of nonaxisymmetric perturbations in a galaxy. BB01 inspected the
two-dimensional ratio (``butterfly") maps of tangential force $F_T$ to
mean radial force $F_{0R}$ to insure that
this maximum was mostly measuring a bar for their 36 galaxies.
Hence, they called the
force ratio maximum $Q_{\rm b}$. A more automated approach
was used by Laurikainen, Salo, \& Rautiainen (2002) and Block
et al. (2002) whereby the maximum force ratio is derived from the
radial variation of the maximum ratio $Q_{\rm T}=|F_{\rm T}/F_{0R}|_{\rm max}$. Depending on the
relative importance of the bar and the spiral, the
maximum $Q_{\rm T}$ could be in the bar or the spiral region. For this reason,
BBK proposed calling
the maximum force ratio $Q_{\rm g}$ to remove any ambiguity
about what it represents.
Since $Q_{\rm g}$ in a barred galaxy could be affected by spiral arm
torques, we cannot use $Q_{\rm g}$ alone to assess whether stronger
bars correlate with stronger spirals. Instead, we need to use the special
technique outlined by BBK, a Fourier-based method that separates
the torques due to a bar from those due to a spiral. However, unlike
estimates of $Q_{\rm g}$, bar/spiral separation is not automated but depends
on an iterative procedure.

In this section, we carry out the BBK
separation procedure on each sample galaxy, and then compare the
results with estimates based on arm-interarm and bar-interbar
contrasts. The galaxy images are deprojected using mainly RC3
(de Vaucouleurs et al. 1991) orientation parameters, with 
revisions for three galaxies based on isophotal ellipse fits (NGC 
6951, 7723) or kinematic parameters (NGC 1530, Regan et al. 1996).
Gravitational potentials are evaluated using the method of
Quillen, Frogel, \& Gonz\'alez (1994), under the assumptions
of a constant near-IR mass-to-light ratio and an exponential
vertical scaleheight $h_z$ estimated from the radial scalelength $h_R$
and the type-dependent ratio $h_R/h_z$ from de Grijs (1998).
A revised lookup table for the vertical dimension (see BBK) was
used that did not involve any gravity softening.

\subsection{Application of the BBK Technique}

In the BBK approach, relative Fourier intensity amplitudes are derived
from the deprojected near-infrared images. In the inner parts of the galaxy,
we assume that the nonaxisymmetric amplitudes are due entirely to the
bar, while past a radius $r_m$ (where $m$ is the Fourier azimuthal index) 
the observed amplitudes are due to a
combination of the bar and the spiral. We assume that the even
relative Fourier amplitudes due to the bar decline past the maximum at $r_m$
in the same manner as they rose to that maximum. A Fourier
image of the bar is constructed and removed from the original
image, to give a spiral plus disk image. Then the $m$=0 Fourier image is
added back to the bar image to give the bar plus disk image. We convert
each of these images into gravitational potentials, and 
derive maps of the ratio of the tangential force to the mean background
radial force, the latter being defined by the $m$=0 Fourier image.
From these ratio maps, we derive the maximum force ratios $Q_{\rm b}$
due to the bar and $Q_{\rm s}$ due to the spiral arms. In practice, we base
our analysis entirely on Fourier-smoothed images using terms up to
$m$=20. BBK fully discuss the pitfalls and uncertainties in this
method of separation, but it is probably the most straightforward
approach one can use for this purpose.

Although BBK illustrated
the technique using a fairly symmetric barred spiral (NGC~6951),
our present sample includes one asymmetric barred spiral, NGC~7741,
that necessitated special attention to odd Fourier amplitudes in
the bar. NGC~7741 is a late-type (SB(s)cd) barred
spiral with strong asymmetry in the bar, a characteristic which
is actually fairly typical of Sd-Sm spirals (de Vaucouleurs
\& Freeman 1972). To deal with it, we took advantage of the fact
that NGC~7741 has no bulge, and chose a center for the Fourier analysis
based on the faintest discernible isophotes of the bar. With
such a center, the odd Fourier terms in the bar go to zero near the
bar ends. In this circumstance, we can define the bar by even and
odd Fourier terms and allow for its asymmetry. We have not accounted
for a slight asymmetry in the bars of two other cases, NGC~972 and
NGC~7479, because these have bright central regions and the effect
is less important.

The success of bar/spiral separation depends on the
quality of the near-infrared images. In our case, the WHT
images are sufficiently well-exposed that noise levels are
minimal in the outer parts of the images. Very bright star-forming
regions were removed from the images to minimize their impact.
These could have a different 
mass-to-light ratio than the surrounding old disk stars, and could
give false torque amplitudes.

Application of the BBK method is not automatic and requires
some iteration for the best choice of the radii $r_m$. Plots of
$I_m/I_0$, the amplitude of the $m$th Fourier component relative
to the $m$=0 component, were initially made for each galaxy and were
evaluated by eye. The most reliable extrapolations were chosen
by examining the resulting bar+disk and spiral+disk images.
Poor choices of extrapolation would sometimes leave artifical
depressions in the spiral+disk image, especially near the
ends of the apparent bar. The final extrapolations used are
those which provided the cleanest-looking separations. BBK
estimated the errors on $Q_{\rm b}$ and $Q_{\rm s}$ due to extrapolation
uncertainties, and showed that a $\pm$10\% uncertainty in the
choice of $r_2$, the radius of the maximum of the dominant
$m$=2 term, can lead to a $\pm$4\% uncertainty in $Q_{\rm b}$ and
$\pm$10\% uncertainty in $Q_{\rm s}$, at least for NGC~6951. We expect
that this is a reasonable estimate of the uncertainty for the other galaxies
in our sample as well (see section 3.2).

We show Fourier extrapolations for the 12 sample galaxies that
required bar/spiral separation in Figure ~\ref{allfourier}.
(The ones that did not require such separation, NGC 908, 1058,
6643, 7217, and 7606, did not show a measurable bar signal.)
In these plots, the curves are the relative intensity amplitudes, and
only the lower order terms are shown (the actual analysis used
terms up to $m$=20 as in BBK). The symbols show the mappings
(interpolations and extrapolations) used for the bar.
Except for NGC~1255, NGC~7723, and NGC~7741, the mappings
are symmetric around a radius $r_m$. For $r < r_m$, the rising
relative amplitudes are used as observed, while for $r > r_m$, the
amplitudes are extrapolated as the exact reverse of the rise.
For NGC~1255, NGC~7723, and NGC~7741, the observed amplitudes
already decline past $r_m$, and in these cases a smaller amount
of extrapolation was needed.

These curves show some of the
different characteristics of the bars in the sample galaxies.
For example, both NGC~1365 and NGC~1433 have bars that can be
extrapolated with relatively flat-topped $m$=2 profiles. Although
the bar is nothing more than a weak oval in NGC~1255, it still
shows a distinct signature in the $m$=2 profile that required
little extrapolation. The bar is stronger in NGC~7723, but it
dominates the amplitudes inside $r$=25$^{\prime\prime}$ and required
little extrapolation past $r_m$. In NGC~7479, the bar and strong
spiral produce a double-humped $m$=2 profile, with one hump corresponding
to the bar and the other hump corresponding to the spiral. NGC~7741
shows an asymmetric $m$=2 profile.

Bar/spiral separation is most reliable when the bar has a relatively constant
position angle in the disk. Figure~\ref{allphases} shows plots of
the $m$=2 Fourier phase $\phi_2$ as a function of radius for the same 12 galaxies
as in Figure~\ref{allfourier}. The solid vertical
lines show the radius, $r(Q_{\rm b})$, of the bar maximum from the force
ratio maps (see Table 2). These show that the assumption of a constant bar phase is
fairly good for the more strongly barred galaxies in the sample, but
is less good for the weakly-barred ones. Nevertheless, the extrapolations
still produce reasonable separations even for the weaker cases.

Figure~\ref{allrotc} shows plots of the rotation curves predicted from
the light distribution for each sample galaxy except NGC~7217 (see Buta et al.
1995 for a detailed study of this galaxy).
These curves are derived from the $m$=0 component of the gravitational
potential as $V=\sqrt{r d\Phi_0/dr}$ and are normalized to the
maximum values. The curves look relatively typical of high luminosity spirals.

Figures ~\ref{allimages} and ~\ref{allratios}
show the analysis images and ratio maps, respectively,
for the 12 sample galaxies requiring bar/spiral separation.
Three images are shown for each galaxy (following BBK):
the $m$=0-20 Fourier sum image, the separated bar+disk image, and the
separated spiral+disk image.
In all cases except NGC~1808, the bar or oval has been rotated to the
approximate horizontal position. For a classical barred spiral like NGC~1365,
Figure~\ref{allimages} shows that the extrapolations in Figure~\ref{allfourier}
produce a fairly
clean separation of the bar and the strong spiral. The ratio maps
for NGC~1365 in Figure~\ref{allratios} show a very regular pattern of alternating
force maxima/minima for this galaxy. In the case of NGC~1433, the
extrapolations in Figure~\ref{allfourier} also produced a clean separation,
but the separated spiral is an oval ring-shaped pattern. Hence, the
spiral shows a butterfly pattern similar to the bar. The remaining
galaxies each show distinctive characteristics in both the images
and ratio maps, but the main point is that the separations appear to
be reasonable representations of the galaxies. In the cases of NGC~1255
and 6814, where the bar is nothing more than an $m$=2 oval, the
four maximum points lie at about $\pm$40$^{\circ}$-45$^{\circ}$ to the bar axis,
while for the more strongly-barred cases, these points lie much closer
to the bar axis. As noted by BB01, higher order terms in the bar
potential cause the maximum points to lie closer to the bar axis.

Several of the more highly inclined galaxies in our sample,
NGC~908, 972, 5033, 6643, and 7606, required bulge/disk
decompositions in order to minimize the effects of bulge
"deprojection stretch." The idea is to model the bulge profile,
subtract off the bulge from the 2D images, deproject the residual
disk light, and then add the bulge back to the deprojected disk
light. Since none of these five galaxies is strongly barred,
one-dimensional decomposition techniques were adequate for our purposes.
We used either the Kent (1986) iterative method or $r^{1\over 4}$
bulge and exponential disk decompositions (e.g., Kormendy 1977).

The bar separation is perhaps less certain for NGC~5033 than for the
other galaxies. The bulge decomposition left some
residual zones of lower intensity around the galaxy minor axis. A small
bar is found in the inner regions, but its ratio map is affected by
the decomposition uncertainties. The ratio map also highlights the
bar-like nature of the spiral in this galaxy.

Of the strongly barred spirals with bright centers, the asymmetry
in NGC~7479's bar left a residual pattern in the spiral
plus disk image. One sees a low intensity region on one side of the
center, and a higher intensity region on the other side.
Ignoring this asymmetry will not affect $Q_{\rm b}$ too
much in this case, since $Q_{\rm b}$ is based on an average of the maximum
points in the four quadrants, but will cause us to underestimate
the scatter in these maximum points.

Our refined procedure for NGC~7741 accounts for the asymmetry
in the bar very well, but leaves a sharp edge in the area
around the bar ends. This has very little impact on $Q_{\rm b}$, but
$Q_{\rm s}$ could have a larger uncertainty because of it.

From the separated ratio maps, we derived plots of the maximum
force ratio, $Q_{\rm T}(r) = |F_{\rm T}/F_{0R}|_{max}$ as a function of radius
$r$ in the galaxy plane. The way these curves are derived is fully
described by BBK. At each radius, we locate the force ratio maximum
in four quadrants and then average the results. Figure~\ref{qtvsr}
shows the curves for 16 of our sample galaxies, including the
ones for which no clear bar signal was detected. In this figure, the radii
are normalized to $r_0(25)$ =  $D_o$/2, where $D_o$ is the Galactic 
extinction-corrected
face-on blue light isophotal diameter at a surface brightness level of
25.0 mag arcsec$^{-2}$ (RC3). The results for NGC~7217 are not shown
since little forcing due either to a bar or a spiral
was detected in this object. Most of the signal seen in the force-ratio map is
likely to be due to bulge deprojection stretch, and has $Q_T$ $\leq$ 0.04 everywhere.
Also, NGC~7217 is the most bulge-dominated
system in our sample (Buta et al. 1995), and our procedure will
not account for this reliably since, like BB01, we have transformed
the light distributions into potentials assuming all components
have the same vertical scaleheight. For our main analysis we have set
$Q_{\rm b}$ $\approx$ $Q_{\rm s}$ = 0 for NGC 7217.

Table 2 summarizes the results of the separations. Four other sample
galaxies, NGC~908, 1058, 6643, and 7606, also had little or no
detectable bar, and we have simply set $Q_{\rm b}$=0 for these.
The weak ovals in NGC~1255 and 6814 were easily separated from their
spirals. The remaining galaxies have a range of bar strengths
up to $Q_{\rm b}$=0.61 in NGC~1530. Owing to a limited field of view,
the spiral parameters for NGC 1808 are more uncertain than the
listed errors imply.

The radii of the maximum relative torques tend to be well inside
the standard isophotal radius, for both bars and spirals. For the
seven objects in Table 2 having $Q_{\rm b}$ $>$ 0.25,
$<r(Q_{\rm b})/r_0>$ = 0.24$\pm$0.09 (s.d.) and $<r(Q_{\rm s})/r_0>$ = 0.42$\pm$0.09
(s.d.). For these same galaxies, $<r(Q_{\rm b})/r(Q_{\rm s})>$ = 0.57$\pm$0.13 (s.d.),
so that the maxima tend to be well-separated on average.

\subsection{Uncertainties}

The main uncertainties in the GTM are due to the uncertainties in the
assumed vertical scaleheight $h_z$, variations in the mass-to-light ratio
(both due to dark matter or stellar population differences),
the adopted orientation parameters
(inclination, line of nodes), the bulge deprojection stretch, the bar thickness,
the sky subtraction, and the galaxy asymmetry. In addition to these,
bar/spiral separation involves uncertainties due to the method of extrapolation.
For the more highly inclined galaxies in the sample, we have minimized
bulge deprojection stretch using bulge/disk decomposition.

BLS show that the uncertainties in
maximum relative torques average about 12\% due to uncertainties in the vertical
scaleheight. A nonconstant vertical scaleheight in the bar could
lead to a further 5\% uncertainty in $Q_{\rm b}$ (Laurikainen \& Salo 2002).
Uncertainties in galaxy inclinations lead to an
inclination-dependence in the uncertainty in $Q_{\rm b}$ or $Q_{\rm s}$.
For an error of $\pm$5$^{\circ}$ in inclination $i$, the error in
maximum relative torques ranges from
4\% for $i\leq$35$^{\circ}$ to 20\% for $i\geq$60$^{\circ}$.
An uncertainty of $\pm$4$^{\circ}$ in line of nodes position angle $\phi$
has less of an impact, ranging from 4\% for $i\leq$35$^{\circ}$ to 11\%
for $i\geq$60$^{\circ}$. Typical sky subtraction uncertainties on the
WHT images could lead to an additional 3.5\% uncertainty in $Q_{\rm b}$ and 5\% uncertainty
in $Q_{\rm s}$ (BBK). Finally, uncertainties in the bar extrapolations and
in the symmetry assumption of the relative Fourier intensity amplitudes
can lead to an additional 4\% uncertainty in $Q_{\rm b}$ and an 11\% uncertainty in $Q_{\rm s}$ (BBK).
Table 2 lists the total uncertainty on each parameter from this analysis.

For NGC 1365, the uncertainty in the orientation parameters may be
larger than we have assumed. Jorsater \& van Moorsel (1995)
analyzed an HI velocity field of NGC 1365 and concluded that circular motions
may be reliable only between radii of 120$^{\prime\prime}$ and
240$^{\prime\prime}$, where the kinematic position angle and inclination
are relatively constant. The kinematic position angle, 220$^{\circ}$,
agrees well with the RC3 value of 212$^{\circ}$. However, the kinematic
inclination of 40$^{\circ}$ significantly disagrees with the nearly
58$^{\circ}$ inclination implied by the RC3 logarithmic
isophotal axis ratio of $logR_{25}$=0.26. Since the bar of NGC 1365
is nearly along the galaxy's minor axis, the derived torque parameters
will be sensitive to this disagreement.

The impact of dark matter on maximum gravitational torques for the OSU bright
galaxy sample was shown by BLS to be fairly small, especially for those
galaxies more luminous than $L_*$=2.5$\times$10$^{10}$ $L_{\odot}$ in the
$B$-band (corresponding to $M_B^o$=$-$20.5). Because the galaxies in the
OSU sample are on average luminous, massive systems, the correction
to $Q_g$ was generally less than 10\% with an average of about 5\%.
Our sample here has similar characteristics, with an
average luminosity of $L_B$=$L_*$. The maximum
relative gravitational torques in both the bar and the spiral regions tend to
lie in the bright inner regions of the galaxies, where the dark halo
contribution is small in such luminous galaxies. Thus, it is likely that
dark matter has only a minimal impact on our results.

The stellar mass-to-light ($M/L$) ratio is a separate issue that could impact
gravitational torque calculations.
Bell \& de Jong (2000) used simple spectrophotometric evolution models
of spiral galaxies to show that stellar $M/L$ variations can, in fact, be
significant, even in the $K$-band. They present simple relations between color
index and the $M/L$ correction for a given passband. However, we cannot make a
reliable deduction from their analysis of the error committed by ignoring
these variations. Reddening can be significant in color index maps, and
would invalidate any $M/L$ corrections from the Bell \& de Jong
relations in some regions, such as bar dust lanes. Also, tests we have made
with $V-K_{\rm s}$ color index maps of NGC 1530 and NGC 7723
indicate that the near-IR spirals are decoupled from the optical spirals
(see section 6) and are not necessarily much bluer than bars, implying that
the mass-to-light ratios of the spirals might not be very
different from those of bars. However, interbar and interarm regions
can be bluer than these features, implying that we could be
underestimating $Q_b$ and $Q_s$ slightly.  Since our sample
is defined by intermediate to late-type spirals, it is likely that $Q_b$
and $Q_s$ are affected by stellar $M/L$ variations in a similar manner for
each galaxy, thus largely preserving any relationship between them.

\section{Correlation Between Bar and Spiral Strength}

\begin{figure}
%\figurenum{7}
\plotone{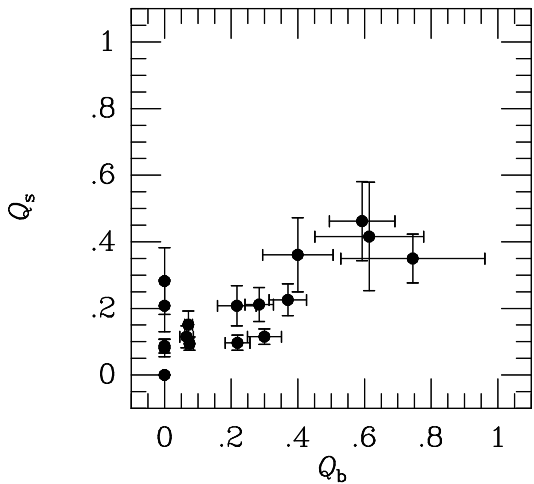}
\caption{Plot of the maximum relative spiral torque, $Q_{\rm s}$, versus the
maximum relative bar torque, $Q_{\rm b}$. Strong bars and spirals are towards
the top and right, respectively.}
\label{qsqb}
\end{figure}

With our estimates of separate bar and spiral strengths, we can
now check how well these two quantities correlate.
Figure~\ref{qsqb} shows $Q_{\rm s}$ versus $Q_{\rm b}$ assuming
constant $M/L$. For $Q_{\rm b}$ $<$ 0.3, there is only a weak
correlation with $Q_{\rm s}$, meaning that spirals as strong as
$Q_{\rm s}$$\approx$0.3 are possible even in the absence of a bar.
However, the strongest spirals in our sample are only associated
with bars having $Q_{\rm b}$ $\geq$ 0.4. The error bars tend to be
large for these stronger cases in part because two of the galaxies
having $Q_{\rm b}$ $>$ 0.4 (NGC~1365 and 1530) have RC3
inclinations close to 60$^{\circ}$ and one (NGC~7741) has
considerable asymmetry in the bar. Figure~\ref{qsqb} also suggests
that among barred galaxies, stronger bars have stronger spirals.
That is, for $Q_{\rm b}\geq 0.2$, $Q_{\rm b}$ and $Q_{\rm s}$ are
linearly correlated.

\section{Other Indicators of Bar and Spiral Strength}

In addition to gravitational torques, the relative importance of
bars and spirals can be assessed
in a variety of different ways. For example, the deprojected ellipticity
of a bar is thought to be a good indicator of bar strength (e.g.,
Martin 1995), based on the theoretical study by Athanassoula (1992).
Abraham \& Merrifield (2000) have refined this idea using an automated analysis
of two-dimensional surface brightness distributions and a rescaled bar ellipticity
parameter called $f_{bar}$ (Whyte et al. 2002). However, bar ellipticity
is an incomplete measure of bar strength since the latter also depends on the total mass
of the bar (Laurikainen, Salo, \& Rautiainen 2002).

Other estimates of bar importance include the bar-interbar
contrast and the amplitude of the $m=2$ Fourier component of the
bar. Similarly, the arm strength can be assessed through the use
of arm-interarm contrasts.
We used the deprojected images to make these estimates for our sample
galaxies. Arm classes (flocculent, multiple arm,
and grand design; Elmegreen \& Elmegreen 1982) were assigned to
the galaxies according to their deprojected $K_{\rm s}$-band
appearance, as indicated in Table 3; some assignments differ from
their original classification in the $B$-band (e.g., Elmegreen \&
Elmegreen 1987). Intensity cuts were made along the major axis of
the discernible bars (using PVECTOR in IRAF) in order to determine
their radial profiles. Bars with constant surface brightness
profiles are labeled ``flat'' in the table, while bars with
decreasing profiles are labeled ``exponential'' (as discussed in
Elmegreen \& Elmegreen 1985). The lengths of the bars were
determined using a program, SPRITE, which fits ellipses to
isophotes via least-squares (see Buta et al. 1999).
Bar length is taken to be the major axis length at maximum
ellipticity, close to where the position angle begins to change.
The maximum ellipticities $\epsilon_b$ and bar radii $r_{bar}$ (relative to the 
extinction-corrected isophotal radius from RC3) are given in Table 3.
In addition to maximum ellipticities and bar radii from the
full images, we also derived these parameters from the {\it
separated bar images}. This is useful for those more strongly
barred spirals where the brightest parts of the arms lie very
close the bar ends. These are listed as $r_{bar}^{\prime}$/$r_0$ and
$\epsilon_b^{\prime}$ in Table 3.

Comparisons between $r(Q_{\rm b})$/$r_0$, $r(Q_{\rm s})$/$r_0$,
and $r_{bar}$/$r_0$ indicate that $r(Q_{\rm s})$ $\approx$ $r_{bar}$
in several of the more strongly barred cases where the inner parts
of the spirals affect the isophotal bar fit. However, 
$r(Q_{\rm s})$ $>$ $r_{bar}^{\prime}$ in all cases.

The deprojected galaxies were transformed into polar images, with gray scale
intensities for radius versus azimuthal angle. The bar-interbar amplitude at
0.7$r_{bar}$ was estimated from intensity profiles parallel and 
perpendicular to the bar. Also, azimuthal intensity cuts with
a width of three pixels were made from the polar images at radii chosen to
be at 0.25 and 0.50 times the standard isophotal radius.
From these profiles, the magnitude
differences between the bars and the interbar regions or between the arms
and the interarm regions were determined for each galaxy, where

$${\rm Contrast (mag)} = 2.5 log (I_{peak} / I_{disk})$$

\noindent
for $I_{peak}$ = bar or arm intensity and $I_{disk}$ = interbar or interarm
intensity. We label the contrast parameters as $A_{bar}$, $A_{0.25}$, and
$A_{0.50}$, respectively. The results are compiled in Table 3.

Fourier transforms for the $m$=2 components of the intensity cuts at 0.7$r_{bar}$ 
were made using the equation

$$F(2)=\sqrt{[\sum{I(\theta) sin(2\theta)}]^2 +[\sum {I(\theta) cos( 2
\theta)}] ^2]} / \sum(I(\theta))$$

\noindent
where $I(\theta)$ is the intensity at each azimuthal angle $\theta$ (ranging
from 0$^{\circ}$ to 360$^{\circ}$),
and the sums are over all angles. These are also tabulated in Table 3.

\begin{figure}
%\figurenum{8}
\plotone{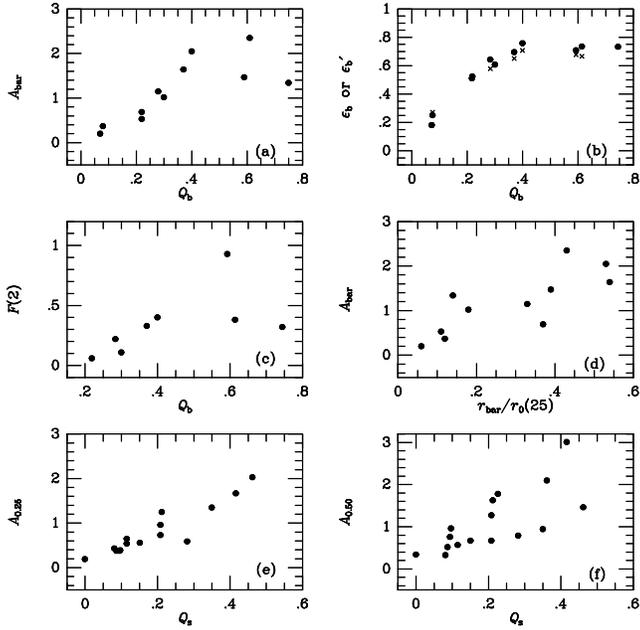}
\caption{Plots of (a) bar-interbar contrast at 0.7$r_{bar}$ versus $Q_{\rm b}$;
(b) bar ellipticity versus $Q_{\rm b}$ (filled circles are $\epsilon_{\rm b}$
for the full images,
while crosses are $\epsilon_{\rm b}^{\prime}$ for the separated bar images); (c) $m$=2 Fourier
transform amplitude versus $Q_{\rm b}$; (d) bar-interbar contrast versus
bar radius relative to the (extinction-corrected)
face-on standard isophotal radius; (e) arm-interarm
contrast at 0.25 times the standard isophotal radius versus $Q_{\rm s}$;
and (f) same as (e), for 0.5 times the standard isophotal radius.}
\label{debbie}
\end{figure}

The results reveal several correlations among these parameters. The
bar-interbar contrast $A_{bar}$ scales in a nearly linear fashion
with $Q_{\rm b}$, but with a considerable scatter at large $A_{bar}$, as shown in Figure~\ref{debbie}a. 
The bar ellipticity also correlates with bar torque,
as shown in Figure~\ref{debbie}b, although the relation is not linear.
This is fully consistent with the results of Laurikainen, Salo, \&
Rautiainen (2002), who derived this same correlation from images of more than
40 galaxies from the Two Micron All-Sky Survey (2MASS). For $Q_{\rm b}$ $<$ 0.4,
the correlation between $\epsilon_b$ and $Q_{\rm b}$ is fairly good, while
for $Q_{\rm b}$ $>$ 0.4, the correlation is weaker in the sense that
small changes in $\epsilon_b$ can correspond to large changes in $Q_{\rm b}$.
The correlation is only slightly improved when $\epsilon_b^{\prime}$ is used.
The Fourier strength parameter $F(2)$ correlates with Q$_b$ (with some
scatter at large torque), as shown
in Figure~\ref{debbie}c. These results indicate that the bar-interbar contrast,
ellipticity, $Q_{\rm b}$, and $m$=2 Fourier component all provide reasonable
measures of bar strength, especially for the weaker bars.

The bar-interbar contrast increases with increasing bar radius, as shown in
Figure~\ref{debbie}d, where $A_{bar}$ is plotted versus the fractional bar 
radius $r_{bar}/r_0$. Furthermore, the bar torque is stronger for flat bars 
(average value of $Q_{\rm b}$ is 0.43$\pm0.14$ for
6 objects) than for exponential bars (0.28$\pm0.16$ for 4 objects). These
results are consistent with
previous $K$-band studies, which indicated that flat bars are both longer and
stronger than exponential bars (Elmegreen \& Elmegreen 1985, Regan \&
Elmegreen 1997). The previous studies also found that earlier spiral types
tend to have flat strong bars and later types tend to have exponential weak
bars, although a correlation with Hubble type is not obvious in the present
small sample. However, correlations of maximum relative gravitational
torques with Hubble type have been established by Laurikainen, Salo, \&
Rautiainen (2002), BLS,
and Laurikainen, Salo, \& Buta (2004). Early-type bars and spirals
tend to be weaker than late-type bars and spirals owing to the dilution
effect of bulges in early-types. In the present sample, the strongest
bar is that in NGC~7741, which also lacks any significant bulge.
Thus, the scatter in the correlation between bar strength and bar length
could be influenced by the importance of the bulge.

The arm-interarm contrast $A_{0.25}$ scales with the arm-interarm
contrast $A_{0.50}$ (not shown) but is slightly weaker, which is consistent with
previous findings that the arm-interarm contrast increases with radius to
about mid-disk (Elmegreen \& Elmegreen 1985). There is a weak correlation
(not shown) between arm strength and Arm Class, with grand design galaxies 
having stronger spirals than flocculent galaxies. The arm-interarm contrast
increases linearly with the maximum relative spiral torque $Q_{\rm s}$, 
as shown in Figure~\ref{debbie}e. The correlation is poorer for $A_{0.50}$.

\section{Dust-Penetrated Classification}

We have applied the dust-penetrated classification scheme of Block \& Puerari
(1999) to each of our sample galaxies.
On the basis of deprojected near-infrared images,
evolved stellar disks may be grouped into three principal
dust penetrated archetypes:
those with tightly wound stellar arms characterized by pitch angles at
$K_{\rm s}$ of $\sim$ 10$^{\circ}$ (the $\alpha$ class),
an intermediate group with pitch angles of $\sim$ 25$^{\circ}$ (the
$\beta$ class) and
thirdly, those with open spirals demarcated by pitch angles at $K_{\rm s}$ of
$\sim$ 40$^{\circ}$ (the $\gamma$ bin).
To take full cognizance of the duality of spiral structure and
decouplings between
gaseous and stellar disks, it has been demonstrated (e.g. Block \&
Wainscoat 1991) that we require {\it two} classification
schemes -- one for the Population I disk, and a separate one for
the Population II disk.  A near-infrared classification scheme can
never {\it replace} an optical one, and vice-versa, because the {\it current}
distribution of old stars strongly affects the {\it current}
distribution of gas in the Population I disk.

For dust-penetrated classification, logarithmic spirals of the form
$r=r_o {\rm exp} (-m \theta /p_{\rm max})$ (see Danver 1942)
are employed in the decomposition.
The amplitude of each Fourier component is given by (Schr\"oder et al.,
1994)

$$\displaystyle A(m,p) = \frac{\Sigma_{i=1}^I \Sigma_{j=1}^J I_{ij}
({\rm ln}\;r,\theta)\; {\rm exp}\;(-i(m \theta + p\; {\rm
ln}\;r))}{\Sigma_{i=1}^I \Sigma_{j=1}^J I_{ij} ({\rm ln}\;r,\theta)}$$

\noindent where $r$ and $\theta$ are polar coordinates, $I({\rm
ln}\;r,\theta)$
is the intensity at position $({\rm ln}\;r, \theta)$, $m$ represents the
number of
arms or modes, and $p$ is the variable associated with the pitch angle
$P$ of a given $m$-mode,
defined by $\displaystyle \tan P = -\frac{m}{p_{\rm max}}$.
We select the pitch angle
of the dominant $m$-mode to define the dust-penetrated pitch angle classes
described above.
Of course, the spiral arms in barred galaxies often depart from a
logarithmic shape. As noted by Block et al. (2001), the arms
may break at a large angle to the bar and then wind back
to the other side, as in a ``pseudoring.''
We minimize the impact of non-constant pitch angle due to rings or pseudorings
angles by excluding from our analysis the bar regions of the galaxies in
question, and by restricting the fits to a limited range in radius. 

To illustrate what the Fourier method is extracting, we show in
Figure~\ref{ngc1530} $m$=2 inverse Fourier transform contours superposed on
our deprojected $K_{\rm s}$ image of NGC 1530. This galaxy has a very open
two-armed spiral breaking from near the ends of its bar. The dominant
$m$-mode is in this region, and the contours are a reasonable representation
of the pitch angle of the arms.

Table 4 summarizes the results of our pitch angle analysis.
The radius ranges used for the fits are listed in column 5, except
for NGC 1365 and 1433 whose classifications are from Buta \& Block (2001).
We continue to find  a ubiquity of low order $m=1$ and $m=2$ modes in this
sample, consistent with earlier studies (e.g. Block \&  Puerari 1999).
The most uncertain classifications are for NGC 1808 and NGC 5033, 
where the field of view limits the part of the main spiral that we can see. 

Under very special circumstances, dominant modes with $m$ greater
than 2 may develop within the modal theory of spiral structure.
Block et al. (1994) and Bertin (1996) hinted that in a {\it
gas-rich} system, some dominant higher-$m$ modes should develop,
and this might also induce some response in the stellar disk, for
the stronger cases. Furthermore, non-linear modes may couple and
again give rise to higher-$m$ structures: $m=2$ and $m=1$ combine
to give $m=3$, $m=2$ and $m=2$ combined to give $m=4$, etc.
(Elmegreen, Elmegreen, \& Montenegro 1992; Block et al., 1994 and
G.~Bertin, private communication). We use the terminology H3 and
H4 for these third and fourth harmonics, to assist with easy
visualization of their evolved disk morphologies.

Finding examples of true three- and four-armed spiral galaxies in the
near-infrared is a great challenge, requiring the investigation of
many galaxies to encounter one or two unambiguous
cases. A good example of an $m=4$ stellar disk in the present study
is NGC~6814, while a previously published example is ESO 566$-$24
(Buta et al. 1998; Rautiainen et al. 2004).

We also note that early Hubble type galaxies (e.g. NGC~972, Sab) can
present very wide open arms in the near-infrared; NGC~972 belongs to the
$\gamma$ class. Three other  examples are NGC~1808 (of RC3 type Sa but
of dust penetrated arm class $\gamma$), NGC~1530 and NGC~1365 (both
Hubble type SBb;
near-infrared arm class $\gamma$). Conversely, it is also possible for
spirals, classified as late type in the optical regime, to have arms
in the near-infrared which are not very wide open (e.g. NGC~908 and
NGC~1058 are of
Hubble type Sc, but both belong to the more tightly wound $\beta$ bin:
see Table 4).

The present study continues to illustrate
decouplings between gaseous and stellar disks and the great advantage
of dust-penetration.  In the optical,
NGC~972 is classified as
flocculent by Elmegreen \& Elmegreen (1987; see also Table 3). The
photograph reproduced by Sandage \&  Bedke (1994)
in panel 148 fully supports the flocculent
designation.
Sandage \& Bedke (1994) note that the optical image of
NGC~972 {\it ``is
dominated by the heavy dust lanes crossing the near-side of the
high-surface-brightness bulge''}. In the
near-infrared, NGC~972 presents a principally two-armed spiral;
the dominant harmonic is $m$=2. Its grand
design, two-armed, evolved stellar disk appears to be completely
decoupled from its fleece-like flocculent Population I gas disk.
While many optically flocculent spirals may still present a flocculent
appearance in the near-infrared (Elmegreen et al. 1999),
decouplings of the two components, when present, are indeed
very striking.

NGC~1808 is famous for its dust lanes which appear to radiate almost
perpendicularly to the major axis. Sandage \& Bedke (1994) note in
their panel 193, that {\it
``the central region of NGC~1808 provides what appears to be direct
evidence of a galactic fountain composed of narrow dust lanes
perpendicular to the plane.''} The near-infrared imaging successfully
penetrates the dust in this SABa spiral, yielding an inner pair of grand
design, wide-open arms (the dominant harmonic in NGC 1808 is $m$=2). In this
case, bar/spiral separation has the added uncertainty that the bar
position angle is not constant (see Figure~\ref{allphases}) and treats
the wide-open spiral as part of the bar.
Also in this case, there is a strong {\it coupling} between the gas and
star-dominated components, although the inner near-infrared morphology
is far more regular. 

We use Table 1 of BB01 to define three gravitational torque
classes: the total torque class based on $Q_{\rm g}$, the bar
torque class based on $Q_{\rm b}$, and the spiral torque class
based on $Q_{\rm s}$. The dust-penetrated type (DP-type), written
in the form ``harmonic class-pitch angle class-bar torque class''
following BB01, is given in the last column of Table 4. Except for
NGC~7741, {\it the highest spiral torque classes all belong to the
highest pitch angle class, $\gamma$}. Six of the galaxies could
not be assigned a DP-type.

\section{Discussion}

The theory of bar-driven spirals is somewhat controversial.
Kormendy \& Norman (1979) showed that of 33 galaxies having
differential rotation and global spiral structure, nearly 80\% are
barred and likely to have their spirals driven by the bars.  The
implication was that bars and spirals would have the same pattern
speeds.  However, Sellwood \& Sparke (1988) showed that in
$n$-body numerical simulations where an initially unstable disk
forms both a bar and a spiral, the spiral has a lower pattern
speed than the bar. Sellwood \& Wilkinson (1993) suggested that
the quadrupole moment of a realistic bar falls off too rapidly to
drive the spirals well beyond the ends of the bar. In this
circumstance, the spiral could be an independent instability, or a
driven response to a resonance interaction between the bar and
spiral. Tagger et al. (1987) and Yuan \& Kuo (1997, 1998) have
shown that resonance interactions between the bar and the spiral
can be an efficient mechanism of wave amplification.

The question of whether spirals and bars corotate can be answered
in some cases by the observed alignment of resonance rings. Inner
rings and pseudorings in SB galaxies tend to be aligned parallel
to bars, while outer rings and pseudorings may be aligned parallel
or perpendicular to bars (Kormendy 1979; Schwarz 1984; Buta 1986,
1995). The morphologies of the various ring types suggest they are
driven by bars: inner rings often have pointy oval shapes,
indicating a connection with the inner 4:1 ultraharmonic resonance
(UHR, Buta \& Combes 1996; Salo et al. 1999), while outer
pseudorings have three distinctive morphologies that suggest a
connection to the outer Lindblad resonance (OLR). These are the
R$_1^{\prime}$, R$_2^{\prime}$, and R$_1$R$_2^{\prime}$
morphologies known as the "OLR subclasses'' (Buta 1985, 1986,
1995; Buta \& Crocker 1991). The first two types were predicted by
Schwarz (1981) from test-particle simulations as being possibly
due to differences in the gas distribution around the OLR, while
Byrd et al. (1994) showed that the three morphologies may be
connected in an evolutionary manner.

Since resonance rings probably develop from the secular evolution
of spiral patterns (Schwarz 1981; Byrd et al. 1994; Rautiainen \&
Salo 2000), the frequent alignments suggest that the spirals that
form the rings are driven by the bars at the same pattern speeds.
Buta \& Combes (1996) further argue that the co-existence of
several rings in the same galaxy, showing shapes and alignments
that are compatible with periodic orbits near resonances (e.g.,
NGC 3081, Buta \& Purcell 1998), speaks against the existence of
several patterns with different pattern speeds. Salo et al. (1999)
show that a model that assumes the bar and spiral co-rotate can
fit the morphology and the velocity field of the early-type ringed
galaxy IC 4214.

The situation is a little different with the central regions of barred
galaxies. Here it is possible for the primary bar to induce enough
mass flow into the central kiloparsec to cause an inner bar instability,
leading to formation of a secondary bar with a different pattern speed
(Pfenniger \& Norman 1990). Support for the idea of a different pattern
speed comes from studies of relative primary bar, secondary bar
position angles (e.g., Buta \& Crocker 1993; Wozniak et al. 1995).

Rautiainen \& Salo (2000, hereafter RS) have reconsidered ring
formation in $n$-body models with dissipatively colliding test
particles. These models show the same features as previous
test-particle models except that with the self-gravitating stellar
disk, additional spiral modes can develop in the same manner as in
Sellwood \& Sparke (1988). RS consider how these modes, which
often have a pattern speed less than that of the bar, affect the
morhology and formation of outer rings and pseudorings. The OLR
subclasses of outer rings and pseudorings still form in these
models, but cyclic changes between the morphologies can occur due
to the influence of the modes with a lower pattern speed. If the
influence of the lower pattern speed modes is significant enough,
RS show that misalignments may be possible. Observed misalignments
between inner and outer rings and bars are rare but do occur (Buta
1995; Buta, Purcell, \& Crocker 1995).

The only prominent ringed galaxies in our sample are NGC 1433 and
NGC 7723. Of these, NGC 1433 shows many features in common with
test-particle simulations, including a highly elongated and
aligned inner ring likely connected with the inner UHR, an
R$_1^{\prime}$ outer pseudoring likely connected with the OLR, a
misaligned nuclear ring, and two secondary spiral arcs, or plumes,
in the vicinity of the outer UHR (Buta et al. 2001). Elmegreen \&
Elmegreen (1985) argue that a bar like that in NGC 1433 could
still be growing through the slow loss of angular momentum to the
stellar spiral. In NGC 7723, deprojection of a blue-light image
indicates that the inner ring is a nearly circular feature from
which a prominent multi-armed spiral pattern emerges. This galaxy
does not fit into available test-particle simulations nearly as
well as does NGC 1433.

We can see in the plots in Figure~\ref{qtvsr} that some of the bar
torques in our sample are weak in the spiral region, as predicted
by Sellwood \& Wilkinson (1993). This is the case for NGC 972,
1255, 5033, and 6814. This may also be true in the Milky Way. Our
Galaxy has a bar about 1.5 kpc long which generates the 3 kpc arm
and perhaps makes the bulge, which is clearly bar-like and about
the right length (see, e.g, Cole \& Weinberg 2002). Corotation is
at $\approx$3 kpc. The spirals outside this radius, which include
the local spiral, the Sgr-Carina spiral, and the Perseus spiral,
may be independent of the bar. This means their pattern speed is
unrelated to the bar pattern or perhaps related in a complex way
through resonant excitation. In either case, the bar strength and
the spiral arm strength should not be simply related. For
resonance excitation, even a weak perturbation can cause a strong
response.

\begin{figure}
%\figurenum{9}
\plotone{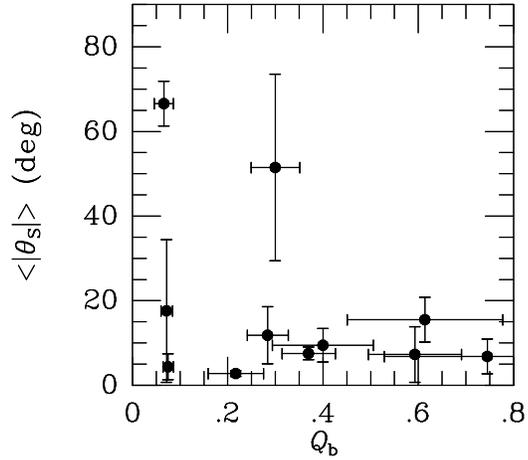}
\caption{Plot of the azimuthal angle of the beginning of the inner spiral,
averaged over two or more arms, versus the bar strength $Q_b$. The error
bars on the angles are average deviations, while those on $Q_b$ are
total mean errors.}
\label{thetaS}
\end{figure}

Another test of the corotation of bars and spirals concerns the
place in azimuth where the inner part of the spiral lies relative
to the bar. If the pattern speeds are different or the bar and
spiral are unrelated, then the distribution of azimuths for these
inner-limit spiral points should be uniform. If all spirals end in
their inner regions at the same angular distance from the end of
the bar, then the bars and spirals are probably related. To
provide a preliminary answer to this question, we examined the
twelve barred galaxies in our sample and visually estimated
$\theta_S$, the angle between the bar and the inner-limit spiral
points, for each prominent spiral arm in each galaxy. Table 3
compiles $<|\theta_S|>$, the average over the prominent arms, and
Figure~\ref{thetaS} shows how this average correlates with $Q_{\rm
b}$. The plot shows that for all but two of the galaxies, the
spirals appear to begin within 20$^{\circ}$ of the bar axis. The
two objects having $<|\theta_S|>$ $>$ 40$^{\circ}$ are NGC 5033
and 7723, the latter having three spiral arms forming part of the
inner ring. Thus, for none of the most strongly barred galaxies in
our sample does the near-IR spiral appear to begin at a large
intermediate angle to the bar axis. However, in blue light, the
situation can be different. For example, both NGC 1530 and 6951
could be interpreted as having serious mismatches between the ends
of the bar and the beginnings of the spiral. On the other hand,
the weak inner pseudorings in these two objects are made from arms
that break from the ends of the bar and wrap around the other
ends.

\section{Conclusions}

From a sample of 17 bright spiral galaxies having a range of bar
and spiral morphologies and Hubble types, we find the following
from $K_{\rm s}$-band images:

\noindent
1. Bars and spirals can be effectively separated using a Fourier-based
image analysis technique (BBK). From this separation, we can quantify the bar and
spiral strengths in terms of tangential-to-radial force ratios. In the sample,
bar strengths $Q_{\rm b}$ range from 0 to 0.75 while spiral strengths $Q_{\rm s}$ range
from 0 to 0.46.

\noindent 2. The spiral strength $Q_{\rm s}$ correlates with
$Q_{\rm b}$ in a nonlinear fashion: spirals form in non-barred
galaxies so $Q_{\rm s}$ is independent of $Q_{\rm b}$ when $Q_{\rm
b}$ $<$ 0.3, but spiral strengths increase linearly with bar
strengths when bars are present, for which $Q_{\rm b}\geq\sim0.3$.
The effect is not an artifact of stellar $M/L$ variations or dark
matter in these high luminosity spirals.  If bars and spirals grow
together in a global disk instability, then the linear relation
between their strengths imply they have about the same growth
rates.

\noindent 3. The bar-interbar contrast $A_{bar}$ and the
deprojected bar ellipticity $\epsilon_b$ correlate with $Q_{\rm
b}$, with a smaller scatter for $\epsilon_b$. The arm-interarm
contrast also correlates with $Q_{\rm s}$, with a better
correlation for the contrast measured at a radius of 0.25$r_{25}$ compared to 0.5$r_{25}$. Strong
bars tend to be longer relative to the galaxy isophotal radius.
This length-strength correlation may be the result of bar
elongation over time as more and more stellar orbits join the bar
potential (see simulations in Combes \& Elmegreen 1993).

\noindent 4. Dust-penetrated classifications indicate that the
strongest bars with the strongest spirals tend to have the most
open spiral arms.  Their pitch angles are 40$^{\circ}$ or more and
their pitch angle class is $\gamma$.  The openness of a spiral arm
depends on the relative size of the Toomre (1964) length, $2\pi
G\Sigma/\kappa^2$ for total disk column density $\Sigma$ and
epicyclic frequency $\kappa$.  This length is approximately the
separation between the arms. If this length is large compared to
the radius, then the arms are very open. The ratio of the Toomre
length to the galaxy radius is approximately the ratio of the disk
mass to the halo mass. Thus open spirals have relatively massive
disks (see also Bertin et al. 1977; Elmegreen \& Elmegreen 1990).
Our observations imply that the most massive disks have the
strongest bars and spirals. This suggests that bars and spirals
form together in a global disk instability.

\noindent 5. The correlations between the maximum relative
gravitational torques of the bars and spirals and between 
a number of other measures of bar and spiral amplitudes, in addition to
the small angular displacements between the ends of the bars and
the inner parts of the spirals, and the alignments of most
resonance rings with their bars, all suggest that the bars and
spirals in most of our sample have shared the same pattern speeds
for cosmological times. In these cases, the bar and spiral
corotation radii are the same. Other cases with more irregular
ring structures or no evident correlations between bar and spiral
strengths or alignments could have their spirals excited either
independently of the bar, or excited at a higher order resonance,
giving the bars and spirals different pattern speeds and no
correlation in structure. The Milky Way is apparently in this
latter category.

These conclusions do not necessarily apply to SB0 galaxies,
which have strong bars and weak (or no) spirals. This
difference from our present sample illustrates the changing
morphology of bar-spiral patterns over time. Our results here
suggest that the growth phase of a strong bar-spiral pattern is
the result of a combined instability having one pattern speed and
one growth rate in a relative massive disk. This growth may be
spontaneous or it may follow an encounter with another galaxy.
After some time, the bar-spiral growth should slow or stop, but it
does this for different reasons in the bar and spiral regions.
Bars stop growing when they saturate to very large strengths,
placing most of the disk stars within the inner 4:1 resonance
inside the bar. The inner stellar population is very hot when this
happens, because the stellar orbits are highly non-circular.
Spirals stop growing when the stellar disk near corotation also
heats up, but for the spirals, the high stellar velocities are
more random than for the bar, removing the spiral pattern.  The
high stellar dispersion also requires a nearly complete conversion
of gas into stars so that gas dissipation is absent and young
stars with low dispersions no longer form. The result is a
relatively strong, squared-off bar with little dust and star
formation structure in the disk and very weak or no spiral arms,
making the SB0 class.

We thank an anonymous referee for helpful comments on this
manuscript. We also thank S. Stedman for assistance during the
WHT observations. RB acknowledges the support of NSF grant AST 0205143
to the University of Alabama. BGE was partially funded by NSF
grant AST-0205097. DLB is indebted to the Anglo American
Chairman's Fund for continued support; in particular, he wishes to
thank M. Keeton, H. Rix and the Board of Trustees. This research
has made use of the NASA/IPAC Extragalactic Database (NED), which
is operated by the Jet Propulsion Laboratory, California Institute
of Technology, under contract with NASA.
The WHT is operated on the island of La Palma by the Isaac Newton
Group in the Spanish Observatorio del Roque de los Muchachos of the
Instituto de Astrof\'\i sica de Canarias.

\clearpage

\centerline{REFERENCES}

\noindent
Abraham, R. G. \& Merrifield, M. R. 2000, \aj, 120, 2835

%\noindent
%Abraham, R. G., Merrifield, M. R., Ellis, R. S., Tanvir, N. R., \&
%Brinchmann, J. 1999, \mnras, 308, 569

\noindent
Athanassoula, E. 1992, \mnras, 259, 328

\noindent
Bell, E. F. \& de Jong, R. 2000, \apj, 550, 212

\noindent Bertin, G., Lau, Y.Y., Lin, C.C., Mark, J.W.K., \&
Sugiyama, L. 1977, Proc. Nat. Acad. Sciences USA, 74, 4726

\noindent
Bertin, G. 1996 in {\it New Extragalactic Perspectives in the New South Africa},
D. L. Block D.L. \& J. M. Greenberg, eds., Kluwer, Dordrecht, p. 227

\noindent
Block, D. L. \& Puerari, I. 1999, \aap, 342, 627

\noindent
Block, D. L. \& Wainscoat, R. J. 1991, Nature, 353, 48

\noindent
Block, D.L. Bertin, G., Stockton, A., Grosbol, P. Moorwood, A. F. M., \&
Peletier, R. F. 1994, A\&A, 288, 365

\noindent
Block, D. L., Puerari, I., Knapen, J. H., Elmegreen, B. G., Buta, R.,
Stedman, S., \& Elmegreen, D. M. 2001, \aap, 375, 761

\noindent
Block, D. L., Puerari, I., Frogel, J. A., Eskridge, P. B., Stockton, A.,
Fuchs, B. 1999, Ap\&SS, 269, 5

\noindent
Block, D. L., Bournaud, F., Combes, F., Puerari, I., \& Buta, R. 2002,
\aap, 394, L35

Buta, R. 1985, Proc. Astr. Soc. Australia, 6, 56

\noindent
Buta, R. 1986, \apjs, 61, 609

\noindent
Buta, R. 1995, \apjs, 96, 39

\noindent
Buta, R. \& Block, D. L. 2001, \apj, 550, 243 (BB01)

\noindent
Buta, R., Block, D. L., \& Knapen, J. H. 2003, \aj, 126, 1148 (BBK)

\noindent
Buta, R. \& Combes, F. 1996, Fund. Cosmic Physics, 17, 95

\noindent
Buta, R. \& Crocker, D. A. 1991, \aj, 102, 1715

\noindent
Buta, R. \& Purcell, G. B. 1998, \aj, 115, 484

\noindent
Buta, R., Purcell, G. B., \& Crocker, D. A. 1995, \aj, 110, 1588

\noindent
Buta, R., Alpert, A., Cobb, M. L., Crocker, D. A., \& Purcell, G. B.
1998, \aj, 116, 1142

\noindent
Buta, R., Laurikainen, E., \& Salo, H. 2004, \aj, 127, 279

\noindent
Buta, R., Purcell, G. B., Cobb, M. L., Crocker, D. A., Rautiainen, P.,
\& Salo, H. 1999, \aj, 117, 778

\noindent
Buta, R., Ryder, S. D., Madsen, G. J., Wesson, K., Crocker, D. A.,
\& Combes, F. 2001, \aj, 121, 225

\noindent
Buta, R., van Driel, W., Braine, J., Combes, F., Wakamatsu, K.,
Sofue, Y., \& Tomita, A. 1995, \apj, 450, 593

\noindent
Byrd, G., Rautiainen, P., Salo, H., Crocker, D. A., \& Buta, R.
1994, \aj, 108, 476

\noindent
Cole, A. A. \& Weinberg, M. D. 2002, \apj, 574, L43

\noindent
Combes, F. \& Sanders, R. H. 1981, \aap, 96, 164

\noindent Combes, F. \& Elmegreen, B.G. 1993, A\&A, 271, 391

\noindent
Danver, C.G. 1942, Lund Obs Ann. vol 10

\noindent
de Grijs, R. 1998, \mnras, 299, 595

\noindent
de Vaucouleurs, G. \& Freeman, K. C. 1972, Vistas in Astr., 14, 163

\noindent de Vaucouleurs, G. et al. 1991, Third Reference Catalog
of Bright Galaxies (New York: Springer) (RC3)

\noindent
Elmegreen, B. G. \& Elmegreen, D. M. 1989, \apj, 342, 677

\noindent
Elmegreen, B. G. \& Elmegreen, D. M. 1985, \apj, 288, 438

\noindent
Elmegreen, D. M. \& Elmegreen, B. G. 1982, \mnras, 201, 1021

\noindent
Elmegreen, D. M. \& Elmegreen, B. G. 1987, \apj, 314, 3

\noindent Elmegreen, B.G., Elmegreen, D.M., \& Montenegro, L.
1992, ApJS, 79, 37

\noindent Elmegreen, D. M. \& Elmegreen, B. G. 1990, ApJ, 364, 412

\noindent Elmegreen, B.G., Wilcots, E., \& Pisano, D.J. 1998, ApJ,
494, L37

\noindent
Elmegreen, D. M., Chromey, F. R., Bissell, B. A., \& Corrado, K. 1999,
\aj, 118, 2618

\noindent
Eskridge, P., Frogel, J. A., Pogge, R. W., et al. 2000, \aj, 119, 536

\noindent
Jogee, S. et al. 2003, in Maps of the Cosmos, IAU Symposium 216, p. 195

\noindent
Jorsater, S. \& van Moorsel, G. 1995, \aj, 110, 2037

\noindent
Kent, S. 1986, \aj, 91, 1301

\noindent
Knapen, J.~H., de Jong, R. S., Stedman, S., \& Bramich, D. M. 2003, \mnras, 344, 527

\noindent
Knapen, J.~H., Shlosman, I., Heller, C. H., Rand, R. J., Beckman, J. E.,
Rozas, M. 2000, \apj, 528, 219

\noindent
Kormendy, J. 1977, \apj, 217, 406

\noindent
Kormendy, J. 1979, \apj, 227, 714

\noindent
Kormendy, J. \& Norman, C. A. 1979, \apj, 233, 539

\noindent
Laurikainen, E., Salo, H., \& Buta, R. 2004a, \apj, in press

\noindent
Laurikainen, E., Salo, H., \& Rautiainen, P. 2002, \mnras, 331, 880

\noindent
Laurikainen, E. \& Salo, H. 2002, \mnras, 337, 1118

\noindent
Laurikainen, E., Salo, H., Buta, R., \& Vasylyev, S. 2004b, in preparation

\noindent
Lindblad, P. A. B., Lindblad, P. O., \& Athanassoula, E. 1996, \aap, 313, 65

\noindent
Martin, P. 1995, \aj, 109, 2428

\noindent
Packham, C., Thompson, K. L., Zurita, A. et al. 2003, \mnras, 345, 395

\noindent
Pfenniger, D. \& Norman, C. A. 1990, \apj, 363, 391

\noindent
Quillen, A. C., Frogel, J. A., \& Gonz\'alez, R. A. 1994, \apj, 437, 162

\noindent
Rautiainen, P. \& Salo, H. 2000, \aap, 362, 465

\noindent
Rautiainen, P., Salo, H., \& Buta, R. 2004, \mnras, in press (astro-ph/0305530)

\noindent
Regan, M. \& Elmegreen, D. M. 1997, \aj, 114, 965

\noindent
Regan, M., Teuben, P. J., Vogel, S. N., \& van der Hulst, T. 1996, \aj,
112, 2549

\noindent
Salo, H., Rautiainen, P., Buta, R., Purcell, G. B., Cobb, M. L., Crocker,
D. A., \& Laurikainen, E. 1999, \aj, 117, 792

\noindent
Sandage, A. \& Bedke, J. S. 1994, The Carnegie Atlas of Galaxies, Carnegie
Inst. of Wash. Publ. No. 638

\noindent
Sanders, R. H. \& Tubbs, A. D. 1980, \apj, 235, 803

\noindent
Schr\"oder, M.F.S, Pastroriza, M.G, Kepler, SO and Puerari,
I. 1994, A\&AS, 108, 41

\noindent
Schwarz, M. P. 1981, \apj, 247, 77

\noindent
Schwarz, M. P. 1984, \aap, 133, 222

\noindent
Sellwood, J. A. \& Sparke, L. S. 1988, \mnras, 231, 25P

\noindent
Sellwood, J. A. \& Wilkinson, A. 1993, Rep. Prog. Phys., 56, 173

%\noindent
%Sheth, K., Regan, M. W., Scoville, N. Z., \& Strubbe, L. E. 2003, \apj, 592, 13

\noindent
Tagger, M., Sygnet, J. F., Athanassoula, E., \& Pellat, R. 1987, \apj, 318, L43

%\noindent
%van den Bergh, S., Abraham, R. G., Whyte, L. F., Merrifield, M. R., Eskridge, P. B.,
%Frogel, J. A., \& Pogge, R. W. 2002, \aj, 123, 2913

\noindent
Whyte, L., Abraham, R. G., Merrifield, M. R., Eskridge, P. B., Frogel,
J. A., \& Pogge, R. W. 2002, \mnras, 336, 1281

\noindent
Wozniak, H., Friedli, D., Martinet, L., Martin, P, \& Bratschi, P. 1995, A\& AS, 111, 115

\noindent Yuan, C., Kuo, C.-L. 1997, ApJ, 486, 750

\noindent Yuan, C., Kuo, C.-L. 1998, ApJ, 497, 689

\clearpage

\begin{deluxetable}{llccccccc}
\tabletypesize{\scriptsize}
\tablewidth{0pc}
\tablecaption{Properties of the Sample Galaxies\tablenotemark{a}}
\tablehead{
\colhead{Galaxy} &
\colhead{Type} &
\colhead{log$D_o$} &
\colhead{log$R_{25}$} &
\colhead{$B_{\rm T}^o$} &
\colhead{$\Delta$} &
\colhead{$M_B^o$} &
\colhead{source} &
\colhead{Exp. Time} \\
\colhead{} &
\colhead{} &
\colhead{} &
\colhead{} &
\colhead{} &
\colhead{(Mpc)} &
\colhead{} &
\colhead{} &
\colhead{(min)} \\
\colhead{1} &
\colhead{2} &
\colhead{3} &
\colhead{4} &
\colhead{5} &
\colhead{6} &
\colhead{7} &
\colhead{8} &
\colhead{9}
}
\startdata
NGC~\phantom{0}908  & SA(s)c     & 1.78 & 0.36 & 10.35 & 17.8 & $-$20.9 & WHT-INGRID & \phantom{0}73 \\
NGC~\phantom{0}972  & Sab        & 1.55 & 0.29 & 11.48 & 21.4 & $-$20.2 & WHT-INGRID & \phantom{0}64 \\
NGC~1058 & SA(rs)c               & 1.51 & 0.03 & 11.55 & \phantom{0}9.1 & $-$18.3 & WHT-INGRID & \phantom{0}62 \\
NGC~1255 & SAB(rs)bc             & 1.62 & 0.20 & 11.26 & 19.9 & $-$20.2 & WHT-INGRID & \phantom{0}60 \\
NGC~1365 & SB(s)b                & 2.05 & 0.26 & \phantom{0}9.93 & 16.9 & $-$21.2 & CTIO-CIRIM & \phantom{0}40 \\
NGC~1433 & (R$^{\prime}$)SB(r)ab & 1.81 & 0.04 & 10.64 & 11.6 & $-$19.7 & CTIO-CIRIM & \phantom{0}30 \\
NGC~1530 & SB(rs)b               & 1.72 & 0.28 & 11.42 & 36.6 & $-$21.4 & WHT-INGRID & \phantom{0}48 \\
NGC~1808 & (R)SAB(s)a            & 1.81 & 0.22 & 10.43 & 10.8 & $-$19.7 & WHT-INGRID & \phantom{0}56 \\
NGC~5033 & SA(s)c                & 2.03 & 0.33 & 10.21 & 18.7 & $-$21.2 & WHT-INGRID & \phantom{0}16 \\
NGC~6643 & SA(rs)c               & 1.60 & 0.30 & 11.14 & 25.5 & $-$20.9 & WHT-INGRID & \phantom{0}63 \\
NGC~6814 & SAB(rs)bc             & 1.54 & 0.03 & 11.32 & 22.8 & $-$20.5 & WHT-INGRID & \phantom{0}60 \\
NGC~6951 & SAB(rs)bc             & 1.68 & 0.08 & 10.71 & 24.1 & $-$21.2 & WHT-INGRID & \phantom{0}60 \\
NGC~7217 & (R)SA(r)ab            & 1.63 & 0.08 & 10.53 & 16.0 & $-$20.5 & WHT-INGRID & \phantom{0}64 \\
NGC~7479 & SB(s)c                & 1.63 & 0.12 & 11.22 & 32.4 & $-$21.3 & WHT-INGRID & \phantom{0}59 \\
NGC~7606 & SA(s)b                & 1.73 & 0.40 & 10.88 & 28.9 & $-$21.4 & WHT-INGRID & \phantom{0}58 \\
NGC~7723 & SB(r)b                & 1.55 & 0.17 & 11.57 & 23.7 & $-$20.3 & WHT-INGRID & \phantom{0}48 \\
NGC~7741 & SB(s)cd               & 1.65 & 0.17 & 11.43 & 12.3 & $-$19.0 & WHT-INGRID & 100\\
\enddata
\tablenotetext{a}{Col. 1: galaxy name; col. 2: de Vaucouleurs revised Hubble type (RC3);
col. 3: log of corrected isophotal diamater (units of 0\rlap{.}$^{\prime}$1) at $\mu_B$
= 25.0 mag arcsec$^{-2}$ (RC3); col. 4: log of iosphotal axis ratio at $\mu_B$ = 25.0
mag arcsec$^{-2}$ (RC3); col. 5: total corrected blue light apparent magnitude (RC3);
col. 6: distance in Mpc (Tully 1988); col. 7: absolute blue light magnitude;
col. 8: source of image (WHT=William Herschel Telescope; CTIO=Cerro Tololo Inter-American
Observatory); col. 9: total on-source exposure time in minutes}
\end{deluxetable}

\begin{deluxetable}{lllllllllll}
\tabletypesize{\scriptsize}
\tablewidth{0pc}
\tablecaption{Derived Torque Parameters\tablenotemark{a}}
\tablehead{
\colhead{Galaxy} &
\colhead{$h_z$} &
\colhead{$Q_{\rm b}$} &
\colhead{$Q_{\rm s}$} &
\colhead{$Q_{\rm g}$} &
\colhead{$r(Q_{\rm b})$} &
\colhead{$r(Q_{\rm s})$} &
\colhead{$r(Q_{\rm g})$} &
\colhead{$r(Q_{\rm b})/r_0$} &
\colhead{$r(Q_{\rm s})/r_0$} &
\colhead{$r(Q_{\rm g})/r_0$} \\
\colhead{1} &
\colhead{2} &
\colhead{3} &
\colhead{4} &
\colhead{5} &
\colhead{6} &
\colhead{7} &
\colhead{8} &
\colhead{9} &
\colhead{10} &
\colhead{11}
}
\startdata
NGC~908 &  379 &  (0.00) &  0.28$\pm$0.10 &  0.28 & \phantom{00}....   &  \phantom{0}73.0 &  \phantom{0}73.0 &  .... &  0.40 &  0.40 \\
NGC~972 &  372 &  0.22$\pm$0.06 &  0.21$\pm$0.06 &  0.24 &   \phantom{00}9.0 &  \phantom{0}29.0 &   \phantom{00}8.0 &  0.08 &  0.27 &  0.08 \\
NGC1058 &  100 &  (0.00) &  0.09$\pm$0.02 &  0.09 &   \phantom{00}.... &  \phantom{0}18.0 &  \phantom{0}18.0 &  .... &  0.19 &  0.19 \\
NGC1255 &  482 &  0.07$\pm$0.01 &  0.15$\pm$0.04 &  0.17 &   \phantom{00}4.5 &  \phantom{0}23.5 &  \phantom{0}76.0 &  0.04 &  0.19 &  0.61 \\
NGC1365 & 1587 &  0.40$\pm$0.11 &  0.36$\pm$0.11 &  0.49 & 118.5 & 178.0 & 164.0 &  0.35 &  0.53 &  0.49 \\
NGC1433 &  581 &  0.37$\pm$0.06 &  0.23$\pm$0.05 &  0.43 &  \phantom{0}68.0 & 106.5 &  \phantom{0}69.0 &  0.35 &  0.55 &  0.36 \\
NGC1530 &  777 &  0.61$\pm$0.16 &  0.42$\pm$0.16 &  0.73 &  \phantom{0}45.0 &  \phantom{0}65.5 &  \phantom{0}49.0 &  0.29 &  0.42 &  0.31 \\
NGC1808 &  600 &  0.22$\pm$0.04 &  0.10$\pm$0.02 &  0.24 &  \phantom{0}70.5 & 118.0\rlap{:} &  \phantom{0}73.5 &  0.36 &  0.61\rlap{:} &  0.38 \\
NGC5033 &  841 &  0.07$\pm$0.02 &  0.12$\pm$0.03 &  0.12 &   \phantom{00}8.5 &  \phantom{0}29.5 &  \phantom{0}29.5 &  0.03 &  0.09 &  0.09 \\
NGC6643 &  383 &  (0.00) &  0.21$\pm$0.08 &  0.21 &   \phantom{00}.... &  \phantom{0}28.0 &  \phantom{0}28.0 &  .... &  0.23 &  0.23  \\
NGC6814 &  375 &  0.07$\pm$0.01 &  0.09$\pm$0.02 &  0.10 &  \phantom{0}10.0 &  \phantom{0}44.5 &  \phantom{0}15.0 &  0.10 &  0.43 &  0.14 \\
NGC6951 &  640 &  0.28$\pm$0.04 &  0.21$\pm$0.06 &  0.34 &  \phantom{0}31.5 &  \phantom{0}57.5 &  \phantom{0}43.0 &  0.22 &  0.40 &  0.30 \\
NGC7217 &  566 &  (0.00) &  (0.00) &  $<$0.04 &   \phantom{00}.... &   \phantom{0}.... &  \phantom{0}....  &  .... &  .... &  .... \\
NGC7479 &  494 &  0.59$\pm$0.10 &  0.46$\pm$0.12 &  0.71 &  \phantom{0}31.0 &  \phantom{0}50.0 &  \phantom{0}45.0 &  0.24 &  0.39 &  0.35 \\
NGC7606 &  898 &  (0.00) &  0.08$\pm$0.03 &  0.08 &   \phantom{00}.... &  \phantom{0}42.5 &  \phantom{0}42.5 &  .... &  0.26 &  0.26 \\
NGC7723 &  508 &  0.30$\pm$0.05 &  0.12$\pm$0.02 &  0.31 &  \phantom{0}16.0 &  \phantom{0}33.0 &  \phantom{0}16.0 &  0.15 &  0.31 &  0.15 \\
NGC7741 &  274 &  0.74$\pm$0.22 &  0.35$\pm$0.07 &  0.77 &  \phantom{0}15.0 &  \phantom{0}46.0 &  \phantom{0}15.0 &  0.11 &  0.34 &  0.11 \\

\enddata
\tablenotetext{a}{Col. 1: galaxy name; col. 2: vertical scaleheight in pc; col. 3: relative
bar torque parameter (bar strength); col. 4: relative spiral torque parameter (spiral strength);
col. 5: gravitational torque parameter (total nonaxisymmetric strength); col. 6: radius
(arcsec) of maximum average relative bar torque; col. 7: radius (arcsec) of maximum average
relative spiral torque; col. 8: radius of maximum average relative total torque; cols.9-11:
same radii as in cols. 6-8, relative to $r_o$ = $D_o$/2, the extinction-corrected de Vaucouleurs
isophotal radius.}
\end{deluxetable}

\begin{deluxetable}{ccccccccccl}
\tabletypesize{\scriptsize}
\tablewidth{0pc}
\tablecaption{Spiral and Bar Contrasts, Dimensions, and Shapes\tablenotemark{a}}
\tablehead{
\colhead{Galaxy} &
\colhead{Arm Class} &
\colhead{$A_{0.25}$} &
\colhead{$A_{0.50}$} &
\colhead{$A_{bar}$} &
\colhead{$r_{bar}$/$r_{0}$} &
\colhead{$\epsilon_b$} &
\colhead{$r_{bar}^{\prime}$/$r_{0}$} &
\colhead{$\epsilon_b^{\prime}$} &
\colhead{$F(2)$} &
\colhead{$<|\theta_{\rm S}|>$} \\
\colhead{1} &
\colhead{2} &
\colhead{3} &
\colhead{4} &
\colhead{5} &
\colhead{6} &
\colhead{7} &
\colhead{8} &
\colhead{9} &
\colhead{10} &
\colhead{11}
}
\startdata
NGC~\phantom{0}908   & M   &    0.59    &       0.79    &    ....   &    ....   &    ....      & .... & .... & .... &..... \\
NGC~\phantom{0}972   & F   &    0.96    &       1.27    &    0.53   &    0.11   &    0.51      & 0.10 & 0.51 & ....  &3:$\pm$1:\\
NGC~1058             & F   &    0.38    &       0.52    &    ....   &    ....   &    ....      & .... & .... & ....  &.....\\
NGC~1255             & M   &    0.56    &       0.67    &    0.20   &    0.06   &    0.18(exp) & 0.06 & 0.18 & ....  &18$\pm$17\\
NGC~1365             & G   &    ....    &       2.10    &    2.05   &    0.53   &    0.76(flat)& 0.50 & 0.71 & 0.40  &9$\pm$4\\
NGC~1433             & G   &    ....    &       1.78    &    1.64   &    0.54   &    0.70(flat)& 0.37 & 0.65 & 0.33  &8$\pm$2\\
NGC~1530             & G   &    1.67    &       3.01    &    2.35   &    0.43   &    0.74(flat)& 0.33 & 0.67 & 0.38  &15$\pm$5\\
NGC~1808             & G   &    0.39    &       0.96    &    0.69   &    0.37   &    0.52(exp?)& 0.37 & 0.52 & 0.06  &.....\\
NGC~5033             & M   &    0.65    &       ....    &    ....   &    ...    &    ....      & .... & .... & ....  &67$\pm$5\\
NGC~6643             & M   &    0.73    &       0.67    &    ....   &    ...    &    ....      & .... & .... & ....  &.....\\
NGC~6814             & M   &    0.38    &       0.76    &    0.37   &    0.12   &    0.25(exp) & 0.16 & 0.27 & ....  &4$\pm$3\\
NGC~6951             & G   &    1.25    &       1.63    &    1.15   &    0.33   &    0.64(flat)& 0.25 & 0.58 & 0.22  &12$\pm$7\\
NGC~7217             & F   &    0.19    &       0.34    &    ....   &    ....   &    ....      & .... & .... & ....  &.....\\
NGC~7479             & G   &    2.03    &       1.46    &    1.47   &    0.39   &    0.71(flat)& 0.23 & 0.68 & 0.93  &7$\pm$7\\
NGC~7606             & G   &    0.43    &       0.33    &    ....   &    ....   &    ....      & .... & .... & ....  &.....\\
NGC~7723             & M   &    0.54    &       0.57    &    1.02   &    0.18   &    0.61(flat)& 0.18 & 0.61 & 0.11  &51$\pm$22\\
NGC~7741             & F   &    1.35    &       0.94    &    1.34   &    0.14   &    0.73(exp) & 0.14 & 0.73 & 0.32  &7$\pm$4\\
\enddata
\tablenotetext{a}{Col. 1: galaxy name; col. 2: spiral Arm Class
(Elmegreen \& Elmegreen 1987, where F=flocculent, M=multi-armed,
G=grand-design); cols. 3,4: the near-IR arm-interarm contrast
at 0.25 and 0.50 times the RC3 isophotal radius; col 5: the near-IR
bar-interbar contrast at 70\% of the deprojected bar radius; col.
6, deprojected bar radius relative to isophotal radius; col 7:
deprojected near-IR ellipticity of the bar (in parentheses,
exp=exponential-profile bar while flat=flat-profile bar); col. 8:
deprojected bar radius relative to the standard (extinction-corrected)
isophotal radius, based on the separated bar+disk image; col. 9: deprojected 
near-IR ellipticity of the bar, based on the same image;
col. 10: Fourier near-IR $m$=2 amplitude;
col. 11: average inner-limit spiral point angles relative to the
bar axis, in degrees. }
\end{deluxetable}

\begin{deluxetable}{cccccccl}
\tabletypesize{\scriptsize}
\tablewidth{0pc}
\tablecaption{Dust-Penetrated Classification\tablenotemark{a}}
\tablehead{
\colhead{Galaxy} &
\colhead{Torque Class} &
\colhead{Bar Class} &
\colhead{Spiral Class} &
\colhead{Radius Range} &
\colhead{$m$} &
\colhead{$|P|$} &
\colhead{DP Type} \\
\colhead{1} &
\colhead{2} &
\colhead{3} &
\colhead{4} &
\colhead{5} &
\colhead{6} &
\colhead{7} &
\colhead{8}
}
\startdata
NGC~908      &      3           &     0        &      3         & 24--60  &    2    & 28.07  &   H2$\beta$0\\
             &                  &              &                &         &    3    & 38.66  &      \\
NGC~972      &      2           &     2        &      2         & \phantom{0}6--19  &    2    & 38.66  &   H2$\gamma$2\\
NGC~1058     &      1           &     0        &      1         &  \phantom{0}6--12  &    2    & 25.20  &   H2$\beta$0\\
NGC~1255     &      2           &     1        &      2         &  \phantom{0}8--27  &    2    & 31.61  &   H2$\beta$1\\
NGC~1365     &      5           &     4        &      4         & .....   &    2    & .....  &   H2$\gamma$4\tablenotemark{b}\\
NGC~1433     &      4           &     4        &      2         & .....   &    2    & .....  &   H2$\alpha$4\tablenotemark{b}\\
NGC~1530     &      7           &     6        &      4         & 29--58  &    2    & 33.69  &   H2$\gamma$7\\
NGC~1808     &      2           &     2        &      1         & 18--48  &    2    & 69.44  &   H2$\gamma$2\\
NGC~5033     &      1           &     1        &      1         & 10--24  &    2    & 28.07  &   H2$\beta$1\\
NGC~6643     &      2           &     0        &      2         &  \phantom{0}7--41  &   ..    & .....  &   ....0\\
NGC~6814     &      1           &     1        &      1         &  \phantom{0}8--36  &    4    & 34.82  &   H4$\gamma$1\\
             &                  &              &                &         &    2    & 11.31  &      \\
NGC~6951     &      3           &     3        &      2         & 24--60  &    2    & 45.00  &   H2$\gamma$3\\
NGC~7217     &      0           &     0        &      0         & .....   &   ..    & .....  &   ....0\\
NGC~7479     &      7           &     6        &      5         & 24--60  &    2    & 33.69  &   H2$\gamma$6\\
NGC~7606     &      1           &     0        &      1         & 17--48  &    2    & 31.61  &   H2$\beta$0\\
NGC~7723     &      3           &     3        &      1         & 10--24  &   ..    & .....  &   ....3\\
NGC~7741     &      8           &     8        &      4         & 22--34  &   ..    & .....  &   ....8\\
\enddata
\tablenotetext{a}{Col. 1: galaxy name; col. 2: maximum relative gravitational
torque class; col. 3: bar torque class; col. 4: spiral torque class; col. 5: radius range (arcsec) for harmonic
and pitch angle fits;
col. 6: multiplicity of dominant spiral harmonic; col. 7,
pitch angle associated with harmonic component $m$; col. 8: dust-penetrated classification (Block \& Puerari 1999;
Buta \& Block 2001).}
\tablenotetext{b}{Harmonic and pitch angle classes from Buta \& Block (2001).}
\end{deluxetable}

\clearpage


%\clearpage

\begin{figure}
\label{allfourier}
%\figurenum{1 (cont.)}
%\plotone{block.fig01.part2.ps}
\caption{Plots of relative Fourier intensity amplitudes as a function of
radius for 12 galaxies where bar/spiral separation was needed. Symbols show the
extrapolations used for our analysis (see text).
For the stronger bar cases, even terms for $m$=2 (solid curve), 4
(dotted curve), and 6 (short dashed curve)
are shown. For NGC~1255 and NGC~6814, only the $m$
= 2 and 4 terms are shown.}
\end{figure}

%\clearpage

\begin{figure}
\label{allphases}
%\figurenum{2 (cont.)}
\caption{Plots of the phase of the $m$=2 Fourier component for the same
12 galaxies as in Figure 1. The solid vertical lines indicate the
radius of the bar maximum from Table 2.}
\end{figure}

%\clearpage

\begin{figure}
\label{allrotc}
%\figurenum{3 (cont.)}
\caption{Predicted axisymmetric rotation curves for 16 of the sample
galaxies, assuming a constant mass-to-light ratio at 2.1$\mu$m.
Each rotation curve is normalized to its maximum $V_m$.}
\end{figure}

%\clearpage

\begin{figure}
%\figurenum{4}
%\plotone{block.fig04.ps}
%\includegraphics[width={17cm}]{block.fig04.ps}
\caption{Separated bar and spiral images for 12 galaxies.
Three images are shown for each galaxy in the following categories:
$m$=0-20 sum (left), bar+disk (middle), and spiral+disk (right).
Each galaxy is identified only in the leftmost of its three
images.}
\label{allimages}
\end{figure}

\begin{figure}
%\figurenum{5}
%\plotone{block.fig05.ps}
%\includegraphics[width={17cm}]{block.fig05.ps}
\caption{Separated bar and spiral force ratio maps for 12 galaxies.
Three ratios maps are shown for each galaxy in the following categories:
$m$=0-20 sum (left), bar+disk (middle), and spiral+disk (right).
Each galaxy is identified only in the leftmost of its three
maps.}
\label{allratios}
\end{figure}
%\clearpage

\begin{figure}
%\figurenum{6 (cont.)}
\label{qtvsr}
\caption{Plots of the average maximum ratio of the tangential force to
the mean radial force as a function of radius normalized to the
extinction-corrected
de Vaucouleurs standard isophotal radius, for 16 of the sample galaxies.
The separate curves refer to the bar (dashed curve), spiral (dotted
curve), and the $m$=0-20 sum image (solid curve).}
\end{figure}

%\clearpage

%\clearpage

%\clearpage

%\clearpage
\begin{figure}
%\figurenum{10}
%\plotone{block.fig10.ps}
%\includegraphics[width={17cm}]{block.fig10.ps}
\caption{Inverse Fourier transform contours of dominant harmonic
superposed on a deprojected $K_{\rm s}$-band image of NGC 1530.}
\label{ngc1530}
\end{figure}

\end{document}